\documentclass[preprint2]{aastex}
\received{}
\accepted{}
\slugcomment{}

\begin{document}
\title{Testing intermediate-age stellar evolution models with VLT photometry of 
LMC clusters. II. Analysis with the Yale models\footnote{Based on observations collected at  the European Southern Observatory,
Chile (ESO N$^o$ 64.L-0385).} 
}

\author{Jong-Hak Woo\altaffilmark{2},
Carme Gallart\altaffilmark{3,4}, 
Pierre Demarque\altaffilmark{2},
Sukyoung Yi\altaffilmark{5},
Manuela Zoccali\altaffilmark{6}}

\begin{abstract}
We present an analysis, using Yale stellar evolution models,
of the color-magnitude diagrams (CMDs) of three intermediate-age 
LMC clusters, namely NGC~2173, SL~556 and NGC~2155, 
obtained with the VLT. 
The main goal of our project is to investigate the amount
of convective core overshoot necessary to reproduce the CMDs of
relatively metal-poor, intermediate age stellar populations, to check
whether the extrapolation that is usually made from solar
metallicity is valid. In the process, we obtained values for the
binary fraction of each cluster, together with refined age estimates. 
Our method involved the comparison of
the observed CMDs with synthetic CMDs computed using various values of
the overshoot parameter and binary fraction. We conclude
that a moderate amount of overshoot and some fraction of binary stars
are essential for reproducing the observed shapes around the turnoff
in the CMD's of all three clusters: unresolved binary stars fill in
the expected core contraction gap, and make a unique sequence
near the gap, which cannot be reproduced by single stars
alone, even with a larger amount of overshoot. We utilize ratios of
the number of stars in different areas around the core contraction gap
to constrain the binary fraction, which is around 10-20\% 
(for primary-to-secondary mass ratio $\ge$ 0.7) in all three clusters. 
Even if binary stars contaminate the core
contraction gap, it is shown that the overshoot parameter can be
inferred from the color dispersion of the stars around the contraction
gap, regardless of the assumed binary fraction. From our overall
analysis such as, shape of isochrones, star counts, color
distribution, and synthetic CMD comparisons, we conclude that overshoot $\sim
20\%$ of the local pressure scale height best reproduces the CMD
properties of all three clusters. The best age estimates
are 1.5, 2.1 and 2.9\,Gyr for NGC~2173, SL~556 and NGC~2155, respectively.
\end{abstract}


\altaffiltext{2}{Department of Astronomy, Yale University, P.O. Box 208101, New Haven, CT 06520-8101; jhwoo@astro.yale.eu, demarque@astro.yale.edu}
\altaffiltext{3}{Andes Prize Fellow, Universidad de Chile and Yale University}
\altaffiltext{4}{Currently: Ram\'on y Cajal Fellow. Instituto de Astrof\'\i sica de
Canarias, 38200 La Laguna, Tenerife, Canary Islands, Spain; carme@iac.es}
\altaffiltext{5}{University of Oxford, Astrophysics, Keble Road, Oxford OX1 3RH, UK; yi@astro.ox.ac.uk}
\altaffiltext{6}{European Southern Observatory, Karl-Schwarzschild-Strasse 2, D-85748 Garching bei M\"{u}nchen, Germany; mzoccali@eso.org}

\section{Introduction}

Stellar evolution theory is routinely tested using Galactic clusters, 
either globular (in general old, and both metal-poor and metal-rich) 
or open (intermediate-age or young, and generally more metal-rich than
globular clusters). These
clusters reflect the particular star formation and chemical enrichment 
history of the Milky Way, and certainly do not represent all possible 
stellar populations. For example, intermediate-age, metal-poor
populations, which are not well represented in the Galaxy, 
are conspicuous in the Magellanic Clouds.
This is an important part of the parameter space,
which is relevant, for example, to the study of stellar populations of
dwarf galaxies in the Local Group, or to that of galaxies at high redshifts,
and therefore, in early evolutionary stages. 

There are uncertainties that are critical to modeling intermediate-mass
stellar models. Most notable of them are the treatments of convection
(e.g., convective core overshoot) and mass loss.
In this paper, we will address them, using intermediate-age
Magellanic Cloud clusters as comparison templates. 

Convective core overshoot (OS) is one of the physical parameters which
must be taken into account during the core hydrogen burning phase of
the evolution of intermediate-mass stars.
Stellar evolution theory predicts that the CMD of young and
intermediate-age star clusters is greatly affected by the amount of
OS. In the early days of stellar structure theory, the
convective core size was determined by the classic Schwarzschild (1906) 
criterion, ignoring the so-called OS, which is due to
the inertial motion of materials beyond the formal core edge.
Initially, the extent of core OS was thought to be negligible
(Saslaw \& Schwarzschild 1965). Subsequent theoretical studies,
however, have emphasized the importance of convective core OS
in stellar evolutionary models (Shaviv \& Salpeter 1973).

Since then, many observational and theoretical studies have been
devoted to investigating the effects of core OS (Prather \& Demarque
1974; Bressan, Chiosi \& Bertelli 1981; Bertelli, Bressan \& Chiosi
1985; Stothers 1991; Meynet, Mermilliod \& Maeder 1993). It is now
generally believed that a moderate amount of core OS is necessary in
order to reproduce the shape of the core contraction gap observed in
CMDs of galactic open clusters (Maeder \& Mermilliod 1981; 
Stothers \& Chin 1991; Carraro et al. 1993; Daniel et al. 1994; 
Demarque, Sarajedini, \& Guo 1994; Kozhurina-Platais
et al. 1997; Rosvick \& VandenBerg 1998),
but the dependence of core OS on stellar mass is still poorly understood
and the subject of on-going investigations (e.g. Ribas, Jordi,
\& Gimenez 2000).

The extent of OS is not currently derivable through direct theoretical
calculations but can be determined empirically by comparing synthetic
CMDs to observed ones, 
with OS in the models often being parameterized as a fraction of the
pressure scale height at the formal edge of the convective core.
In the CMDs of intermediate-age star clusters ($2 \lesssim$ Age $\lesssim
5$\,Gyr), which have a distinctive gap near the turnoff (TO) caused by the
rapid contraction of a hydrogen-exhausted convective core, the TO
topology can be used to evaluate the extent of core OS
(Demarque et al. 1994; Kozhurina-Platais et al. 1997).
In the case of younger clusters, the importance of core OS can be
estimated from the observed luminosity function of main sequence (MS)
stars (Chiosi et al. 1989; Barmina, Girardi \& Chiosi 2002) and from the ratio
of main sequence to post-main sequence stars. Studies of detached
eclipsing binaries can also provide empirical estimates of OS by
direct comparison with stellar tracks in the log $T_{\rm eff}$ - log $g$
plane (Schroder, Pols, \& Eggleton 1997; Guinan et al. 2000; 
Ribas et al. 2000). However, the number of binary systems
available for OS studies is very limited.

Many galactic open cluster studies have suggested a moderate amount of
OS, i.e. $\sim 20 \%$ of the local pressure scale height at the edge
of a convective core. However, this has been tested only for near
solar metallicity clusters in the Galaxy. So far, the presence of OS
in metal-poor clusters in the Magellanic Clouds has only been
tested in a limited sample of young (age $\le$ 500 Myr) clusters
(Lattanzio et al. 1991; Vallenari et al. 1991, 1994; Stothers \& Chin
1992; Brocato, Castellani \& Piersimoni 1994; Chiosi et al. 1995;
Testa et al. 1999; Keller, Da Costa \& Bessell 2001; Barmina et al. 2002). 
For Magellanic Clouds intermediate-age
clusters, the necessary observations are more challenging due to the
fainter magnitude of the TO, and HST or 8-m class telescopes
under excellent seeing conditions are required. 

The characteristics of mass loss during the RGB phase are still unclear
in many aspects. Among the early papers on the subject, that of Reimers
(1975) was most influential and his mass loss formula 
has been widely used to date: 

\begin{equation}
\dot{M} = -4 \times 10^{-13} \eta \frac{L}{gR}
\end{equation}
where L, g, and R are luminosity, gravity and radius, respectively.
The mass loss efficiency parameter $\eta$, which was later introduced,
has been reported to vary from 0.25 to 2 -- 3 for metal-rich stars
\cite{d86,kr78,r81}.
In contrast, the estimated range of $\eta$ for metal-poor stars appeared 
to be narrow, as most studies on metal-poor stars suggest
$\eta$ = 0.3 -- 0.7 \cite{am82,ma82,r81,ldz94},
and these values successfully reproduce the horizontal branch morphology 
of the old Galactic globular clusters \cite{yi99}. 
While the application of Reimers' formula for low mass stars predicts that 
the total amount of mass loss gets smaller as mass increases, it is
not clearly demonstrated how the amount of mass loss depends on 
the mass of the stars in intermediate-age clusters
(Tripicco, Dorman \& Bell 1993; Liebert, Saffer \& Green 1994;
Carraro et al. 1996). 

In this context, we started a project to investigate several aspects
in the evolution of intermediate-age, metal-poor stellar populations,
in particular, the amount of OS and the mass loss
characteristics. To reach this goal, we observed three
intermediate-age clusters using the Very Large Telescope (VLT)
 under excellent seeing conditions. The project and the observations 
are presented in Gallart et al. (2002; Paper I). 
In the current paper, we will compare the 
observations with the recent set of Yonsei-Yale ($Y^{2}$) stellar evolutionary 
models, and with models specifically calculated for this project using 
the same main input physics (but differing in the OS parameter),
to constrain the amount of OS and the mass loss.

The three clusters under the study seem to present the signature of a core
contraction gap in their TO region, which is predicted by the models as well
for this age range.
A possible obstacle in studies of OS using the shape of the core
contraction gap is binary contamination. 
Castellani, Degl'Innocenti \& Marconi (1999)
pointed out that the contraction gap enhanced by core OS can be filled
in by unresolved binary stars.  In fact, some intermediate-age
clusters, which are expected to have a core contraction gap, show a
continuous sequence rather than a distinctive gap (Carraro et
al. 1994; Paper I). Although this could be considered as a limiting factor
for the topology method, we will show that it is possible, using
synthetic CMDs, to put some constraints on the
binary star population. It may be also expected that an increase of
binary fraction in synthetic CMD simulations tends to decrease the value of OS
required for a good fit.
Testa et al. (1999) have claimed that a
no-OS model with $\sim 30\%$ of unresolved binaries allows a
fair fit to the observed MS luminosity function of NGC~1866 in the
LMC, while Barmina et al. (2002), using a revised analysis of the same
data, have concluded that the inclusion of unresolved binaries would
not alter the need for OS (see also for other clusters, 
Vallenari et al. 1992; Carraro et al. 1994) 

In this paper, we present a short description of the observational 
data (\S~2).
Stellar models including OS treatment and synthetic CMD simulations 
are described in section 3. Detailed analysis on the steps of deriving
ages, binary fraction, the amount of OS and mass loss is described
in section 4, followed by a short discussion (\S~5) and conclusions (\S~6).

\section{THE DATA}

The three LMC clusters NGC~2173, SL~556 and NGC~2155 were observed
with FORS1 at VLT-UT1, in service mode, through the filters $V$ and $R$,
under excellent seeing conditions. PSF fitting photometry was obtained
with the suite of programs DAOPHOTII/ALLSTAR/ALLFRAME (Stetson 1987,
1994), and calibrated using data of the same clusters obtained under
photometric conditions using the MOSAIC camera at the CTIO-4m
telescope.  We estimate an accuracy of the zero points of the
photometric transformation of $\pm$ 0.005. An assessment of the completeness 
and error characteristics of the data was performed through artificial star
tests. Finally, we carefully subtracted the LMC field stars in the cluster 
field in a statistical fashion. All these aspects of the data treatment are
discussed in detail in Paper I. In the following sections we will just use the
photometrically calibrated, field-subtracted CMDs, and the results of
the artificial star tests for completeness and errors.

\section{MODELS}

\subsection{Stellar tracks \& isochrones}

We have constructed stellar tracks of metallicity $Z=0.004$ and $0.007$ with the
Yale stellar evolution code YREC (Guenther et al. 1992), 
ignoring the effects of rotation.
All input physics and model construction assumptions are the same as
in the $Y^{2}$ Isochrones (Yi et al. 2001) except for
the OS treatment.  Three sets of tracks have been generated
depending on the amount of core OS. The OS parameter is
defined as a fraction of the pressure scale height at the edge of a
convective core and prescribes the radial size of the OS
region beyond the classical core convection boundary. We used 10, 20
and 30\% of the local pressure scale height as the amount of
OS, corresponding to ``OS parameter" values of 0.1, 0.2 and 0.3, respectively.

\begin{figure}
\plotone{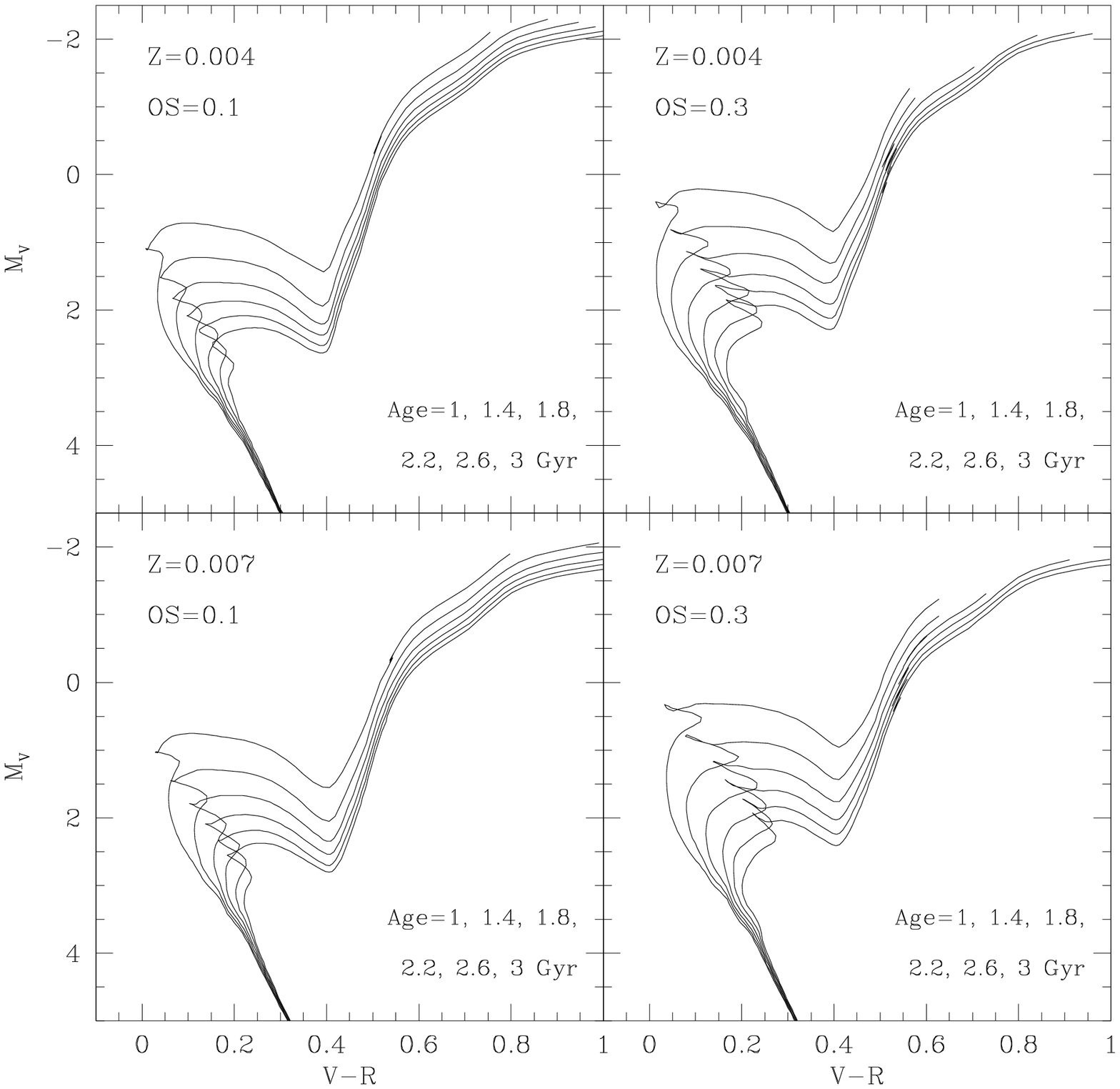}
\caption{Isochrones with OS=0.1 and 0.3 for Z = 0.004 and 0.007, respectively.
Isochrones between 1 -- 3\,Gyr are plotted for two metallicities.
For each given age, isochrones with larger OS have brighter TOs and subgiant 
branches.
}
\label{osiso}
\end{figure}

From these tracks, we have constructed isochrones between 1 and 3\,Gyr,
with an interval in age of 0.1\,Gyr. Color transformations are based on
Lejeune, Cuisinier \& Buser (1998), following the $Y^{2}$ scheme. 
It is generally found that a
convective core develops when the stellar mass is larger than $\sim
1.1$\,$M_{\odot}$, a result known to depend on chemical composition.
For $Y^{2}$ isochrones, Yi et al. (2001) have adopted OS = 0.2 for
young isochrones ($<$ 3\,Gyr) and OS = 0.0 for old ones ($\geq
3$\,Gyr) based on the observational studies of approximately solar
abundance populations. 
Their isochrones have a 1\,Gyr age grid for ages greater than 1\,Gyr
and do not provide the isochrones of ages greater than 2\,Gyr but smaller than 3\,Gyr. This is inconvenient to our study because
our sample clusters appear to be in this age range.
Besides, the $Y^{2}$ Isochrones group does not recommend any
simplistic isochrone interpolation for the age range of 2 to 3\,Gyr
because it is in this age range that the transition between
convective core and radiative core in the MS TO stars occurs
(see also Woo \& Demarque 2001).
For this reason, we have computed a new $Y^{2}$ Isochrones
in which the OS parameter has been kept constant for ages between 1 and 3\,Gyr. 
Figure~\ref{osiso} illustrates the isochrones of two different OS parameters. 
Models with
larger OS show brighter TOs and subgiant branches for each
given age, implying that an age estimate for an observed CMD would be
larger when using isochrones with larger OS.

\subsection{Synthetic CMD code}

Stellar mass is the fundamental parameter that determines the
evolutionary path of each star in the CMD. By randomly distributing
stellar mass with an initial mass function (IMF), and tracing each
star's evolutionary stage at a given time, one can construct synthetic
CMDs for stellar systems with single age and metallicity. 
These simple stellar populations are relatively easy to model due to the
simplifying assumption of instantaneous star formation rate and
homogeneous chemical composition.

Several kinds of uncertainties still remain in computing synthetic
CMDs for simple stellar populations.  The first kind, which affects
each star's location on the CMD, comes from uncertainties in stellar
evolutionary theory and stellar atmosphere models.  Secondly,
uncertainties in the number of stars in each location on the CMD come
from the combinations of uncertainties in the evolutionary time scale of
each stage, the slope of the IMF, and stochastic effects caused by
small number statistics.  Many CMD studies have aimed at testing
physical parameters in stellar evolutionary theory by comparing
theoretical shape and star counts with the observations while
minimizing uncertainties from other sources. Depending on the nature
of the targets and on the methods used in the analysis, these
uncertainties can be minimized with rigorous effort. For example,
plausible stochastic effects in number ratio analysis can be avoided
by repeating the simulations and using large enough areas in the CMD
to count stars in observed and synthetic CMDs.

A new CMD synthesis code has been created, including binary simulation,
based on previous works (Woo 1998; Yi et al. 1999). 
For MS to RGB stars, Salpeter's IMF with x = 1.35 has been used.
For synthetic horizontal-branch construction, we have used the
core helium-burning stellar models of Yi, Demarque \& Kim (1997) 
with Lee, Demarque \& Zinn (1990)'s prescription for mass distribution.
Mass loss along the RGB is formally calculated with Reimers' (1975) formula,
and subtracted at the RGB tip to derive the mass of core 
helium burning stars. However, for this particular study, 
the amount of mass loss is empirically determined to reproduce 
the observed luminosity of the red clump.

Unresolved binary stars make a distinct sequence in the observed CMDs.
It is critical to include binary stars for a detailed study of the
CMD. We define the binary fraction (BF) as the number ratio
of unresolved binary stars to the total number of stars in the CMD.
The binary stars in the CMD are simulated in the following way.  Once
stellar mass is distributed according to the stellar IMF, some stars
are randomly selected to be given a secondary star. The mass of the
secondary star $M_{se}$ is assigned using the critical mass ratio,
$q_{c}$ that sets the minimum mass ratio between secondary and
primary:

\begin{equation}
M_{se}=(RAN*(1-q_{c})+q_{c})*M_{pr}
\end{equation}
where $M_{pr}$ is the primary star mass, and RAN is a random number
with a flat distribution between 0 and 1. 
Such a flat distribution for the mass function of the
secondary stars and a typical value of $q_{c}$=0.7 is adopted
following Elson et al. (1998; see also Hurley \& Tout 1999). 
Secondary stars with masses
$M_{se}\le0.7M_{pr}$ would form a sequence indistinguishable
from that of single stars. The BF that we will quote
throughout the paper will be, therefore, the fraction of binary stars
with $M_{se}\ge0.7M_{pr}$, and a larger {\it total} BF should be expected.

Once the mass of the secondary star is determined, then
the magnitudes $M_{V}$(pr) and $M_{V}$(se) of the primary and
secondary, respectively, are combined to calculate the magnitude
$M_{V}$(BI) for the unresolved binary system:

\begin{equation}
M_{V}(BI)=-2.5* log~[10^{-0.4*M_{V}(pr)} +10^{-0.4*M_{V}(se)}]
\end{equation}

\subsection{Error \& completeness simulation}
The simulation of the observational errors (photometric shifts and
completeness) in the synthetic CMDs has been performed on a
star-by-star basis, using an empirical approach without any assumption
on the nature of the errors or their propagation. The whole process
has been performed in a way similar to that described in Aparicio \&
Gallart (1995) and Gallart et al. (1996a,b), and it is based on the
results of the artificial star tests discussed in Paper I.

We have derived two tables from the artificial star tests. In the first
one, the completeness information for each V, R magnitude interval is
recorded. In the second table, for each artificial star that was
recovered under the same quality criteria as the cluster data, the
input and output magnitudes are listed. From this information, the
simulation process for photometric errors and completeness in the
synthetic CMD is the following: for each star in the synthetic CMD,
with magnitudes $V_s$, $R_s$, we first use the completeness
information in the corresponding magnitude interval to determine the
probability of the star having been lost in the photometry, and keep
it or remove it from the list of synthetic stars according to that
probability.  In a second step, for each synthetic star that has been
kept, we select all artificial stars within a given distance in color
and magnitude from the synthetic star. One of these stars is then
picked up at random, and its $\delta V$ and $\delta R$, calculated as
the differences between the recovered and injected magnitudes for that
star, are added to the magnitudes $V_s$, $R_s$ in order to simulate
the observational errors.

\begin{figure}
\plotone{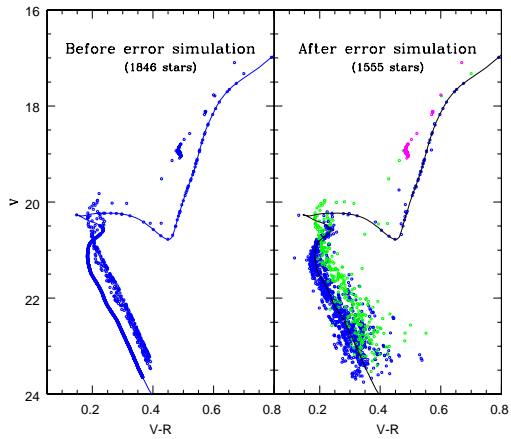}
\caption{Synthetic CMDs before and after the error simulations. 
Left: Single stars are distributed on top of an isochrone.
Unresolved binaries are randomly chosen from single stars,
and dispersed according to the various mass ratio, q ($0.7 \leq q \leq 1$).
Right: Each star is scattered with a photometric error that is randomly
chosen from the artificial stars errors table. More than 10\% of the stars are
lost due to the completeness simulation.
}
\label{errorsim}
\end{figure}

In Figure~\ref{errorsim}, synthetic CMDs before and after error
simulations are compared as an example.  In the left panel, single
stars are placed on top of a given isochrone while the binary sequence
has a dispersion according to the range of the mass ratio q ($q_{c}
\leq q \leq 1$).  The error simulated CMD in the right panel shows many
scattered stars in the CMD due to the photometric errors especially at
faint magnitude levels.  The result of the completeness simulation is
that, globally, 10 to 15\% of the total input stars are lost in the
error simulated CMD.

\section{Analysis: overshoot and mass loss}

In this section, we compare the CMDs of the three LMC clusters, 
NGC~2173, SL~556 and NGC~2155, with isochrones and synthetic CMDs, in 
order to test the predictions of stellar evolution theory under different
sets of assumed parameters regarding OS and mass loss. 
This will allow us not only to derive the cluster parameters
(ages, metallicity, distance, and reddening) but also to constrain
the input parameters in stellar model calculations (e.g., OS and mass loss).

\subsection{The amount of overshoot in low-metallicity, intermediate-age 
populations}

Cluster CMDs provide a powerful tool for deriving the OS parameter
empirically. We have applied the following strategy on our sample clusters.
First, we used the color difference $\Delta$(V-R) between 
the base of the RGB and
the TO to constrain the age range for each cluster under
each OS assumption (Sarajedini \& Demarque 1990). 
This has been complemented with visual isochrone
fitting within that age range, to obtain a single best fitting age for
each OS parameter. Second, since the presence of unresolved binary
stars substantially affects the morphology of the TO and core
contraction gap in the CMD, we attempt to find the BF 
using the ratio of the number of stars in two boxes around
and below the core contraction gap. The last step takes
advantage of the different spread in color near the core contraction
gap depending on the assumed OS, to single out the best OS, and
consequently, the best age for the cluster. 
Binary stars have substantial effects on the morphology of the TO
and the core contraction gap, however,
we will show that they do not greatly affect the color
distribution of the stars, and therefore, the final conclusion on the
best fitting OS parameter is almost independent of the assumed BF.
These steps are explained in detail in the following three
sections.

\subsubsection{Isochrone solutions}

The major effect of core OS to stellar evolution is to supply hydrogen
fuel from the OS-induced mixing region to the core, which lengthens 
the hydrogen core burning phase. This results in a brighter TO and subgiant
branch since for a given age, more massive stars are
going through the core contraction phase. Consequently, if the
amount of OS assumed in the models is larger, an older age will be
derived for a given observed CMD. Therefore, the OS parameter
cannot be uniquely determined without an accurate age constraint and
vice versa.  Due to this degeneracy, we determined the best age for
each OS parameter in the following steps.

\begin{figure}
\plotone{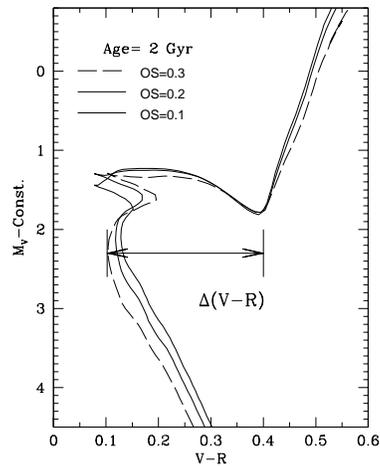}
\caption{$\Delta$ (V-R) dependence on the OS parameter.
We compare the 2\,Gyr isochrones of 3 different OS parameters, which are
vertically shifted to clearly demonstrate the color difference.
The color difference between the base of the RGB and the TO increases
as a function of OS since stellar mass at the TO
increases with larger OS for a given age.
}
\label{delvr}
\end{figure}

As the TO gets redder with increasing age, the color difference,
$\Delta$(V-R) between the base of the RGB and the TO decreases. For
a given age, a larger OS implies an increase of $\Delta$(V-R) since
more massive stars are located near the TO (Figure~\ref{delvr}).
Thus, for each OS parameter, an age can be estimated from the observed
$\Delta$(V-R) value.  However, due to the uncertainties in determining
the locations of the base of RGB and the TO in the observed CMD, this
step only gives a plausible range of age for each OS. A typical error,
0.02 in $\Delta$(V-R) gives age uncertainty $\sim$ 0.3\,Gyr.

Within the age range, we have performed isochrone fitting to find
the best fitting age. In this step, two further parameters are
required: distance modulus and interstellar reddening.
The distance modulus of the LMC is usually taken to be
18.5 $\pm$ 0.1, which is consistent with the recent compilation of
Benedict et al. (2002). However, there is still some disagreement 
on the LMC distance modulus among experts in the field, 
and values ranging
from 18.3 to 18.7 have been advanced. In addition, the LMC has a
large angular distribution of clusters in the sky, among which the
distance modulus may differ. For example, if we consider that the line
of sight depth of the LMC is of the order of the LMC bar size, 3
degree ($\sim$ 2.6 kpc), the distance modulus can easily differ by up
to 0.1 magnitude.  Currently, the distance to individual clusters is
unknown.

A mean reddening of E(B-V)$\simeq$0.1 [E(V-R)= 0.08] or larger
is tipically found for the LMC (Harris, Zaritsky, \& Thompson 1997, 
and references therein). Most of these
studies, however, focus on the central part of the LMC, or on regions
containing young stars. Therefore, presumably, a larger amount
of dust is present in these regions than in more external LMC regions, 
as is the case of the clusters in our study. 
Schlegel, Finkbeiner, \& Davis (1998) calculate typical reddenings 
toward the LMC from the median dust emission in surrounding anuli, 
and find E(B-V)=0.075 (E(V-R)=0.06)),
which should be close to the reddening found in the outer part of the
LMC where internal extiction may be very low. 
No estimates of individual reddenings toward the clusters in our
study are available. 

Given these uncertainties, we performed isochrone
fitting with limited freedom of distance modulus and reddening.  We
did not use a fiducial line obtained from the CMD in the fitting
process since a fiducial line is difficult to determine around the
convective core contraction gap, and binary stars can influence
the fiducial line position. Since field star subtraction is not
perfect, we rather put more weight on the isochrone shape near the
TO and upper MS, which are relatively well populated.

\begin{figure}
\plotone{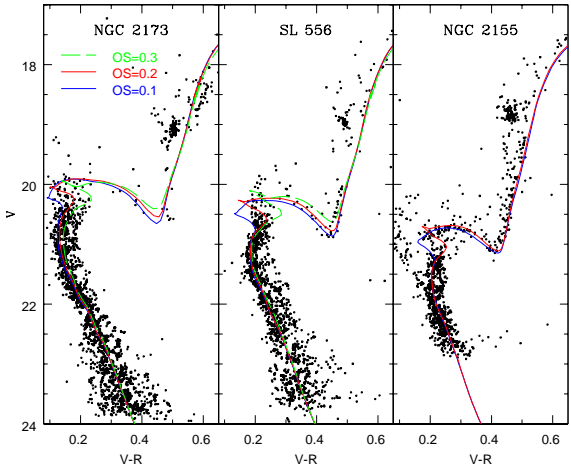}
\caption{Isochrone solutions for each cluster. Isochrones for each 
overshoot-age pair (see Table 1) are plotted on top of the observed CMD.
Due to the uncertainties in field star subtractions, much weight is put
on the populous region of the CMD in the isochrone fitting process.
For example, it is hard to define SGB and RGB on the observed CMD of NGC~2173 since
there are only a few subgiant stars (some of them are binaries and field stars)
and many stars on the upper RGB seem to be field stars.
The isochrone shapes of all three solutions are very similar except
the core contraction gap.
}
\label{ISOSOL}
\end{figure}

With isochrone fitting, we single out the cluster age for each OS
parameter, which is also complemented by comparing the observed CMD with
the synthetic CMD simulated with each set of age and OS (Table 1).
The true metallicities of our sample clusters are still poorly known
but often suggested to be in the range Z=0.004 -- 0.007 (Paper 1).
We prefer to use a single value of metallicity for all three clusters
in order to investigate systematic effects.
Metallicity Z=0.007 seems to be inadequate as a representative value
because the Z=0.007 isochrones do not yield reasonable matches 
(MS and RGB location, TO shape, and core contraction gap) to
all three cluster CMDs due to the redder RGB. 
Therefore, we exclude the Z=0.007 solutions
and only use Z=0.004 solutions for further analysis.
However, it should be noted that NGC~2173 might be more metal-poor
than Z=0.004 (Paper 1).
For NGC~2155, OS=0.3 isochrones produce very poor matches to the
observed CMDs and thus are excluded in the further analysis of this study.
The final isochrone solutions are plotted with the observed CMD of each cluster
(Figure~\ref{ISOSOL}). Note that the shapes of all three isochrones 
for each cluster are very similar except for the core contraction gap.

\subsubsection{Unresolved binary stars}

{\bf i) The effect of binaries and overshoot on the shape of the gap}

The effects of unresolved binary stars require special attention in
cluster CMD studies since the binary sequence modifies the
distribution of stars from the expected single star sequence.
This contamination effect on the TO topology is more
profound to intermediate-age clusters than to much younger
or much older clusters since the core contraction gap is prominent 
only in intermediate-age clusters.
We have performed CMD synthesis including binary stars in order to
investigate how they affect the distribution of stars around
the convective core contraction gap. For this test, we selected
isochrone solutions of SL~556 with various BFs (0 to
30\%).  

\begin{figure}
\plotone{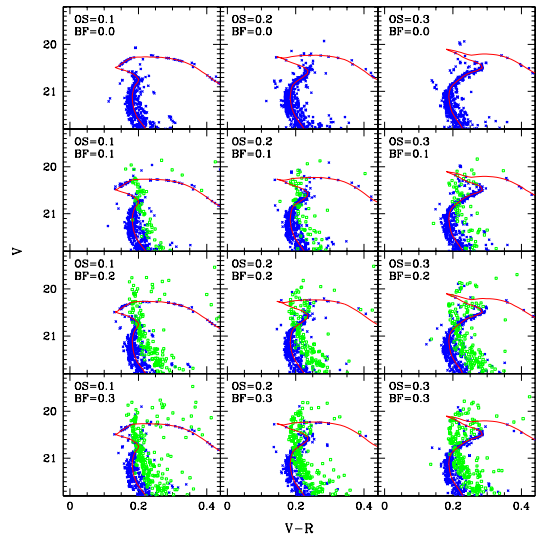}
\caption{The effect of overshoot and binary fraction.
Various shapes of the core contraction gap are presented 
in synthetic CMD simulations
with isochrone solutions of SL~556. Single stars (crosses) and binaries
(squares) have distinct shapes around the gap. 
As OS increases, the MS gets redder and longer due to the delayed core
contraction.
Stars above the core contraction gap cannot be reproduced by single stars with even
larger amount of OS.
For a given OS parameter, more stars fill in the gap with increasing
BF and make a distinctive round shape, which resembles the shape of
single stars near the TO.
}
\label{bf}
\end{figure}

Figure~\ref{bf} shows how the shape of the CMD around the gap
is affected by binaries and OS. 
For a given BF, larger OS
results in a longer MS extending further toward the red due to the
delayed core contraction phase. Consequently, the color distribution becomes
much broader around the gap. Also, since the evolutionary time scale
after core contraction phase is much faster with larger OS, the subgiant
branch is less populous with increasing OS.
For a given OS parameter, more stars are located around the gap with
increasing BF. Binary stars make a rounded sequence, which resembles the
shape of the single star sequence near the TO, which is $\approx$
0.75 magnitude fainter. 
The stars above the gap are most likely binaries
since single stars cannot be located above the gap even with much
larger OS.

Binaries fill in the core contraction gap, which causes visual confusion
near the TO and makes the derivations of age and OS parameter difficult.
However, their sequence can still be differentiated from that of single stars
using a detailed TO topology comparison.
Furthermore, the number of stars above the core contraction gap, 
which are most likely unresolved
binaries according to our analysis, can constrain the BF.

{\bf ii) Constraining the binary fraction}

The binary frequency in a cluster CMD can be determined by
artificial star tests using a binary sequence parallel to the MS
(Rubenstein \& Bailyn 1997). However, this method is time-consuming
and still has a large uncertainty in estimating the BF.
For our clusters, which are of intermediate-age and expected
to display a core contraction gap, the distribution of stars around
the core contraction gap can be utilized for estimating the BF.

\begin{figure}
\plotone{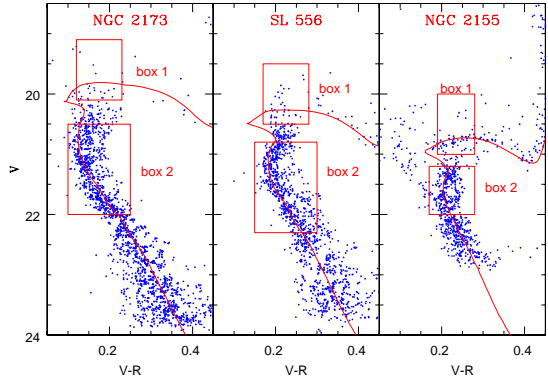}
\caption{The location of boxes for testing binary fraction.
Two boxes are shown in each cluster's CMD. From the binary simulation,
the location of the binary star box (Box 1) is carefully selected
in order to include most likely binaries. The lower box (Box 2) is selected to
count MS stars with some binaries. We have used boxes of the same size,
except for NGC~2155, for which the limiting magnitude is much brighter.
}
\label{box}
\end{figure}

We estimate the BF using the number ratio of the stars above the core 
contraction gap to the stars in the upper MS,
respectively counted in two boxes in the CMD (Figure~\ref{box}). The
location of Box 1 is carefully selected to contain mostly binary
stars and some subgiant stars (although for the oldest cluster 
NGC~2155 the number of subgiant stars increases). 
As mentioned in the previous section, we have compared a synthetic CMD
without binaries to a CMD with 30\% binaries, and determined the
best location for the binary star box (Box~1). The location of
Box~2 is selected to cover the TO region including a large fraction
of MS stars and some binaries. 
If the total number of stars is fixed, a larger value of BF
naturally results in an increase in number of stars in Box 1,
while the number of stars in Box 2 is not significantly affected by the BF.
Therefore, the
number ratio of Box 1 to Box 2 can be a good indicator of the BF in
the observed CMD.

\begin{figure}
\plotone{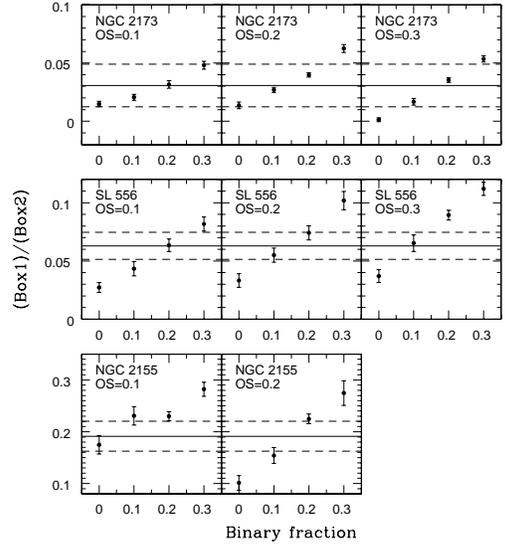}
\caption{The number ratio of Box 1 to Box 2 for each isochrone solution.
The average number ratio of Box 1 to Box 2 is estimated from
10 synthetic simulations with various BFs (circle).
Dispersion from the mean is denoted with an error bar. 
As BF increases, the number ratio increases since more binaries are located
in Box 1. The horizontal lines show the number ratio (solid line) 
from the observed CMD and the errors (dashed line) estimated from the decontamination process.
It is shown that 10 -- 20 \% of binaries (with $q \ge 0.7$) matches the
number ratio of stars above the core contraction gap to stars in the upper
MS.
}
\label{bfsim}
\end{figure}

The number ratio of stars in these two boxes for each cluster is shown
in Figure~\ref{bfsim} with a solid line. The uncertainties of the number
ratios in the observed CMDs are estimated from the field star 
decontamination process and shown with dashed lines.
In spite of the same box
size, the observed number ratio of SL~556 is much greater than that of
NGC~2173, which indicates a larger BF in SL~556. This
result is consistent with the presence of a clear binary sequence
parallel to the MS in the CMD of SL~556.  In case of
NGC~2155, a direct comparison with the other clusters is not possible since
a Box 2 of much smaller size has been used due to the brighter limiting
magnitude.

We have performed 10 Monte Carlo simulations for each given BF and
averaged the number of stars in each box. Dots in
Figure~\ref{bfsim} show the mean number ratio of the two boxes in the
synthetic CMDs, with error bars indicating 1 $\sigma$ dispersion. 
Simulations show a general trend of the binary effect
on the number ratio, which increases with increasing BF as expected.
According to our simulations, the number ratio varies slightly with
different box locations and sizes, however, the trend appears to remain the
same if we keep the same boxes in the observed CMD and the synthetic CMD.
A comparison between the number ratios from the empirical data
and from the simulations suggests 10-20\% of unresolved binaries
(with $q\ge 0.7$, the BF could be larger for any $q$) are required 
for all three clusters. 
These estimates are similar to those of Elson et al. (1998) for 
the LMC cluster NGC~1818 and of Rubenstein \& Bailyn (1997) for the 
Galactic globular cluster NGC~6752.

\subsubsection{Overshoot from the color distribution around the core contraction gap}

\begin{figure}
\plotone{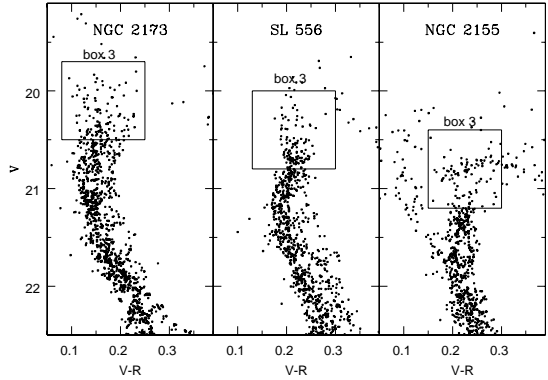}
\caption{The location of Box 3 near the core contraction gap.
A box covering most stars near the core contraction gap is represented
in the CMD of each cluster. The color distribution of stars in this box
for observed and synthetic CMDs will be analyzed to discriminate for the best
OS parameter.
}
\label{box2}
\end{figure}

As mentioned in \S~ 3.1, the empirical TO morphology may constrain
the amount of OS and find the most probable isochrone solution in
Table 1 and Figure~\ref{ISOSOL}. For a quantitative analysis, we have drawn
Box 3 around the core contraction gap and analyzed the color
distribution of stars in the box. The location of the box for each
cluster, which contains most stars in core contraction phase, is shown
in Figure~\ref{box2}.

\begin{figure}
\plotone{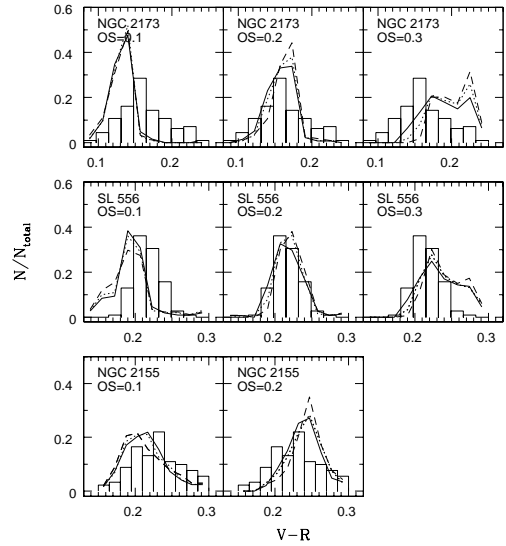}
\caption{Color distribution of stars around the core contraction gap.
The color distribution of the stars in Box 3 from synthetic simulations 
(lines) are compared with that of observation (histogram).
For each OS parameter, 3 different BF values are used (BF=0: solid, 
BF=0.1: dotted, BF=0.2: dashed line). Note that the effect of BF on
the color distribution is insignificant.
In contrast, larger OS tends
to make a broader color distribution. The median color is also shifted
to a redder color with increasing OS. Comparing with the observed color
distribution, the best OS parameters are determined.
}
\label{hooksim}
\end{figure}

For all synthetic CMD models with each isochrone solution (OS and
age), the color distributions of stars in Box 3 are plotted in
Figure~\ref{hooksim}. Each panel contains the distribution of stars
from the observed CMD (histogram) and from simulations with various values
of BF (10\%-30\%), which are suggested in the previous section. 
Note that once scaled to the total number of stars in
Box 3, the color distribution is almost independent of BF, while OS
affects the color distribution significantly. In general, the
distribution is broadened and shifted to the red with increasing OS.

The best fits for NGC~2173 and SL~556 are models with OS=0.2, as
clearly indicated by the median of the distribution.
The various BF does not significantly affect the color distribution
and, thus, OS estimation. The best OS value for NGC~2155, which is 
older than the other clusters, seems to lie between 0.1 and 0.2.
We conclude that for all 3 LMC clusters the OS parameter is
around 0.2. The fact that 
OS seems to be between 0.1 and 0.2 for NGC~2155 may indicate that OS
slightly decreases with cluster age. A larger number of clusters in this 
age range would be necessary to confirm the trend.

\subsection{Mass loss and the location of the red clump}

The distinct feature of the red clump (RC) stars in the CMDs of
intermediate-age clusters deserves considerable attention for its
potential use as a distance indicator. 
However, it can be useful only when its characteristics
(i.e., luminosity dependence on age and metallicity) are properly understood
(Girardi \& Salaris 2001; Grocholski \& Sarajedini 2002).

An age dependence of the RC luminosity is naturally expected
for the younger clusters with TO mass $\gtrsim 1.8
M_{\odot}$, for which helium ignites in a non-degenerate core,
since the core mass of helium burning stars increases as the age of
a cluster is decreasing.
In contrast, in the case of stars undergoing the He-flash, 
the degenerate core mass is nearly constant and independent of age,
but the envelope mass can play a role in controlling the RC luminosity 
since the energy output from the hydrogen burning shell is decreasing with 
a smaller envelope mass.
Thus, if mass loss during the RGB phase depends on age, 
there can be an age dependence in the RC luminosity of intermediate-age clusters.

\begin{figure}
\plotone{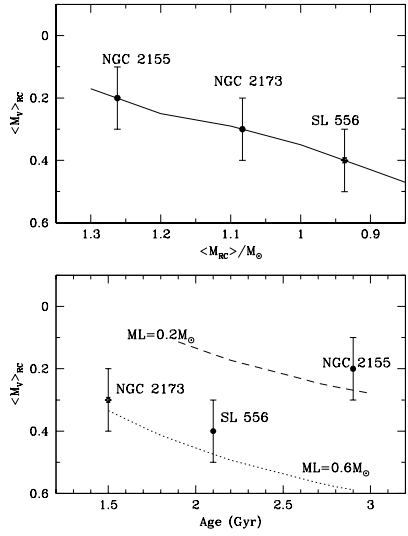}
\caption{Top: relation between the median RC luminosity and the mean RC mass.
Median magnitudes of RC stars are calculated
from simulated CMDs with various mean RC stellar masses (solid line).
Using this relation, the mean RC mass of each cluster is determined 
from the observed RC magnitude. Note that the oldest cluster, NGC~2155
has the brightest RC luminosity.
Bottom: Median RC magnitude of model predictions assuming fixed mass loss, 0.2 $M_{\odot}$ (dashed line) and 0.6 $M_{\odot}$ (dotted line) compared with that of 
observation (points). The age of all three clusters are adopted 
from the OS=0.2 isochrone solution.}
\label{RC}
\end{figure}

In this section, we derive and discuss mass loss for each cluster 
using an empirically obtained relation between RC luminosity and RC mass.
First, we have performed synthetic CMD simulations, using He-burning tracks 
(Yi et al. 1997), and measured the RC luminosity, $M_{V}$, 
as a function of the mean RC mass (Figure~\ref{RC} top panel). 
It is shown that the median RC magnitude increases with larger RC mass,
which is a generic feature of He-burning
stellar tracks at this mass range 
(Vandenberg 1985; Seidel, Demarque \& Weinberg 1987).
Second, RC mass for each observed CMD is empirically derived 
from this luminosity--mass relation (points in Figure~\ref{RC} top panel). 
The observed RC magnitude is converted to $M_{V}$ with the distance modulus,
and a typical error (0.1 mag.) is assumed considering the uncertainties in
distance modulus. 
It is interesting to note that NGC~2173 and SL~556 have a fainter RC 
magnitude than the older cluster NGC~2155, which is also shown in Figure~\ref{ISOSOL}, 
and consequently a smaller mean RC mass is inferred.
Third, the amount of mass loss along the RGB has been estimated by
subtracting the mean RC mass from the RGB tip mass of each isochrone.
The mass loss errors are calculated from the assumed uncertainties
in the RC magnitude. 

We note that the estimated mass loss
value for NGC~2155 is similar to that obtained using mass loss efficiency,
$\eta$= 0.5 - 0.7 in Reimers' formula, which successfully reproduces
the horizontal branch morphology in old globular clusters \cite{yi99}.
However, the mass loss derived for SL~556 is too large to be
compatible with $\eta$= 0.5 - 0.7 and may indicate a much higher mass 
loss efficiency in the context of Reimers' formula.
To illustrate this, we derived RC luminosity, $M_{V}$, for clusters with various
age assuming a fixed mass loss value, 0.2 and 0.6\,$M_{\odot}$, 
respectively (Figure~\ref{RC} bottom panel). 

The origin of such a large estimate of mass loss in SSL~556 and 
disagreement with $\eta$=0.5-0.7 remains unclear to us 
and may have an important implication on the uncertainty/variability
of the mass loss efficiency even in similar environments. 
The validity of mass loss estimates are subject to the 
uncertainties in distance modulus, age and metallicity
although the color difference between the
TO and the RGB is clear enough to indicate a reasonably accurate age
estimate and the assumed metallicity provides a good match on the RGB slope.
Further investigations are essential to draw more confident conclusions,
especially with a better distance modulus constraint.

In the case of NGC~2173, it is unclear whether the RC stars
experienced He-flash since the RGB tip mass of this cluster lies near the
critical mass at which stars develop a degenerate core, although it is
known that the critical mass becomes smaller with increasing OS 
(Chiosi et al. 1989). 
If not, then the lower RC luminosity of NGC~2173
may result from the fact that RC luminosity decreases as
RGB tip mass becomes much larger than 1.5 $M_{\odot}$
(see Girardi, Mermilliod \& Carraro 2000).
However, this cannot be the case for SL~556 and NGC~2155 
since their RGB tip mass is similar or less than $\sim 1.5$ $M_{\odot}$.

\section{Discussion: Age gap among LMC clusters}
Jensen, Mould \& Reid (1988) first noticed a gap in the age distribution of
LMC clusters, with none of them with ages between 4 and 10\,Gyr.
This age gap would be present also in metallicity, with younger LMC
clusters having [Fe/H]$\simeq -0.7$ and older ones having
[Fe/H]$\simeq -2.0$ (Olszewski et al. 1991). Geisler et al. (1997)
performed a survey of candidate old clusters and confirm the
existence of a gap in the LMC cluster formation, which they situate
similarly between 3 and 8\,Gyr ago. Shortly after, Sarajedini (1998)
claimed to have found three 4\,Gyr old LMC clusters, namely NGC~2155, 
SL~556 and NGC~2121, which would be in the age gap. With deeper observations,
Rich, Shara \& Zurek (2001) have estimated the age of these clusters to be
3.2\,Gyr old, and claimed that they are probably the first clusters to
have formed in the younger group of the LMC clusters. Our age
estimate of NGC~2155 is slightly younger (by $\sim 0.3$\,Gyr) but
consistent with that of Rich et al. (2001), supporting the extension of
the LMC age gap up to 3\,Gyr on its intermediate-age side.

\section{Conclusion}

In this paper, we have performed detailed synthetic CMD simulations on
the effect of OS and binary stars, compared with observed CMDs
of three LMC clusters, NGC~2173, SL~556, and NGC~2155.  Unresolved
binary stars fill in the core contraction gap and make a unique
sequence near the gap, which cannot be reproduced by single
stars alone, even with a larger OS amount.  We utilize the
number ratios of stars around the core contraction gap 
to derive a first order estimate of
the BF. Based on our star count analysis, all three clusters 
have a BF of 10\% to 20\% (with $q \ge 0.7$). 

Using the color distribution of stars around the core contraction gap,
which does not significantly depend on the BF, we
find that the best OS parameter for each cluster is close to
20\% of the local pressure scale height. 

With the best solutions of the BF and OS from star count and color
distribution analysis, we present synthetic CMDs for each cluster with
the observed CMD in Figure 11, 12, and 13, respectively. 

From the overall analysis, such as isochrone fitting, star counts,
color distributions, and synthetic CMD comparison, we conclude that a
moderate amount of OS, $\sim 20\%$ of the local pressure scale height,
is essential to reproduce the observed shape around the core
contraction gap in the CMDs of all three clusters, which are more
metal-poor than the intermediate-age open clusters observed in the
Galaxy, implying OS does not depend on
metallicity at least for this metallicity range.
With OS=0.2, our best age estimates are 1.5, 2.1 and 2.9\,Gyr 
for NGC~2173, SL~556 and NGC~2155, respectively. 

We constrainted mass loss along the RGB phase from the observed
RC luminosity of each cluster, utilizing the RC 
luminosity -- mass relation derived from synthetic CMDs. 
We found that the mas loss estimate for NGC~2155 is consistent with 
the typical mass loss parameterization that works for the Galactic globular 
clusters. However, the SL~556 data, 
and perhaps the data for the other young cluster
NGC~2173 as well, indicate a substantially larger amount of mass loss 
than suggested by Reimers' formula. This may indicate once again 
the complexity of mass loss processes.

\acknowledgements 
This research is part of a Joint Project
between Universidad de Chile and Yale University, partially funded
by the Fundaci\'on Andes. Our data was collected as part of an ESO 
Service Mode run. 
C.G. acknowledges partial support from Chilean CONICYT through
FONDECYT grant number 1990638. 
This research has been
supported in part by NASA grant NAG5-8406 (P.D.). 
We thank H. Yong for his contribution in the early stage of this research.

\begin{figure}
\plotone{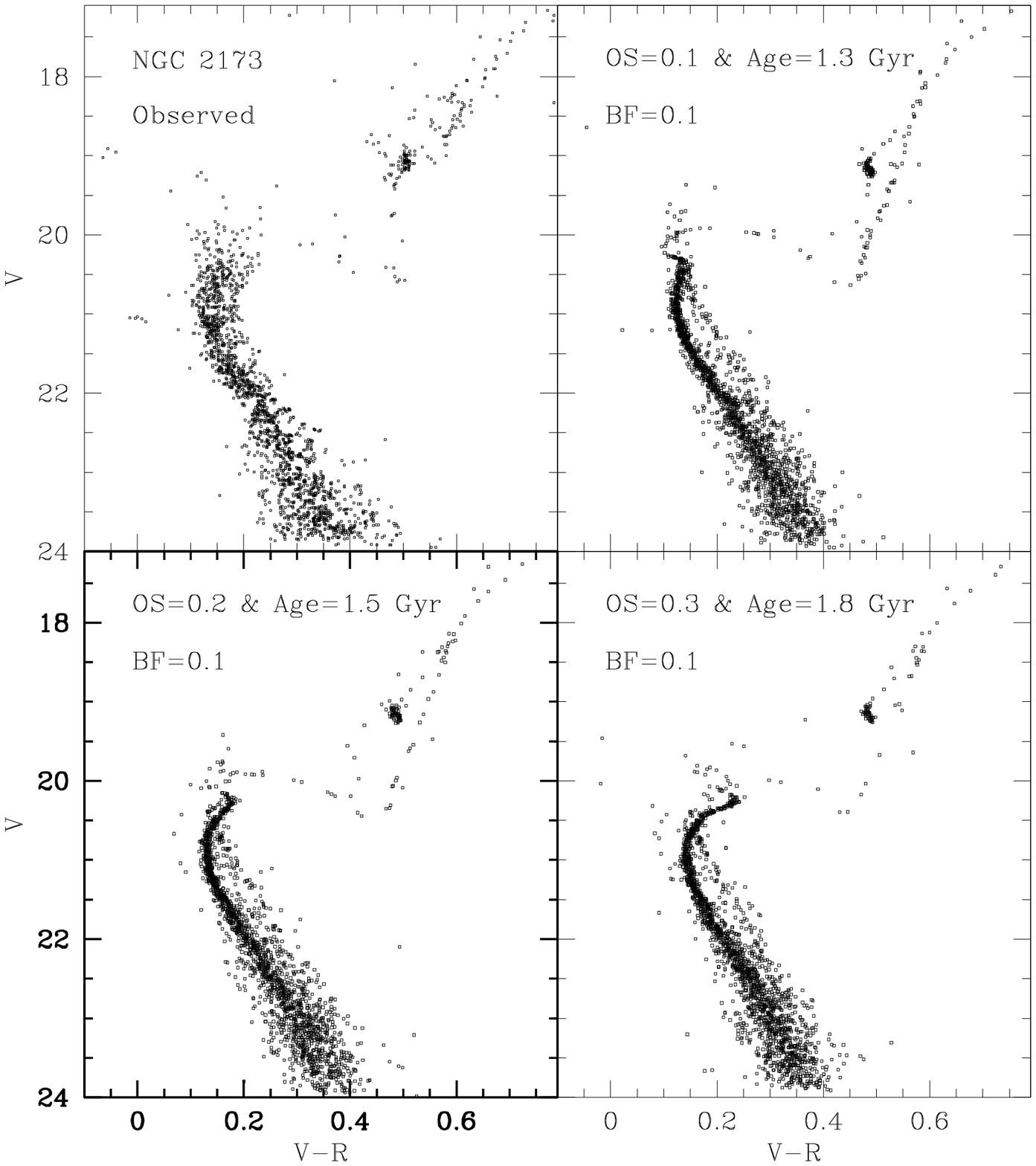}
\caption{Synthetic CMD of NGC~2173.
Each synthetic CMD for a given pair of OS and BF is compared with the
observed CMD. The synthetic CMD with the best solution (OS = 0.2)
is emphasized with a thick box. 
}
\label{syn1}
\end{figure}

\begin{figure}
\plotone{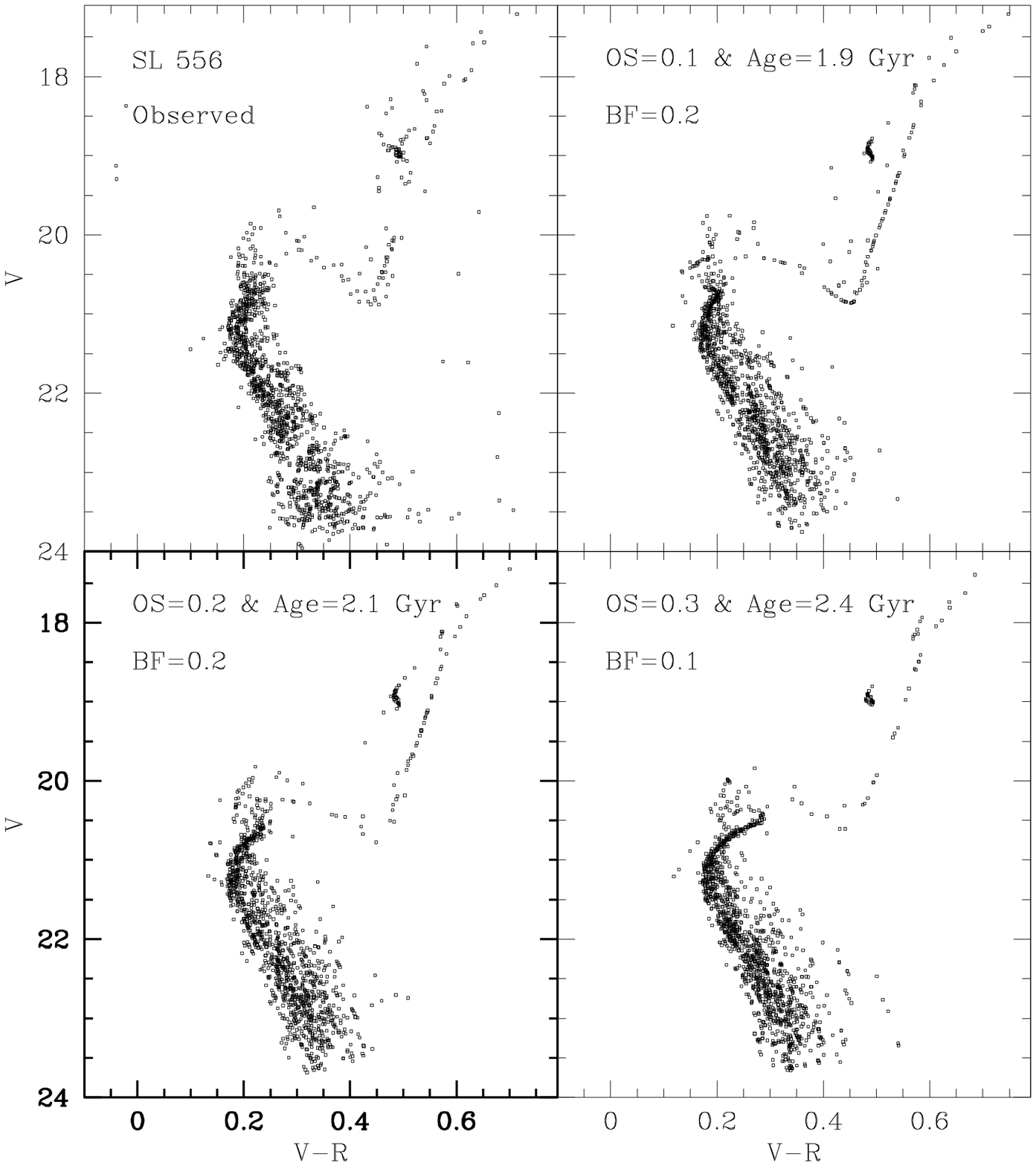}
\caption{Synthetic CMD of SL~556.
Each synthetic CMD for a given pair of OS and BF is compared with the
observed CMD. The synthetic CMD with the best solution (OS = 0.2)
is emphasized with a thick box.
}
\label{syn2}   
\end{figure}

\begin{figure}
\plotone{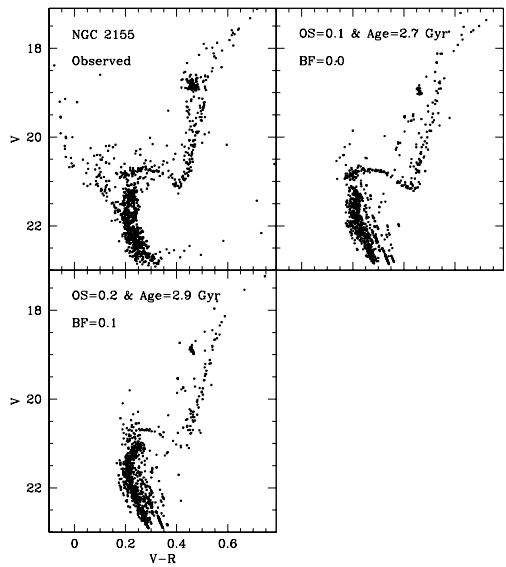}
\caption{Synthetic CMD of NGC~2155.
Each synthetic CMD for a given pair of OS and BF is compared with
the observed CMD. On this plot, it is difficult to discriminate
between OS=0.1 and 0.2.
}
\label{syn3}   
\end{figure}

\begin{table}
\begin{center}
\caption[2]{Isochrone solutions for each cluster}
\begin{tabular}{lcrrrrr}
\hline
Object & Z & Overshoot & Age (Gyr) & $M_{RGB}$ & E(V-R) & $(m-M)_{o}$ \\
\hline
NGC~2173 & 0.004 & 0.1 & 1.3 & 1.751 & 0.06 & 18.5 \\
         &       & 0.2 & 1.5 & 1.695 & 0.06 & 18.5 \\
         &       & 0.3 & 1.8 & 1.630 & 0.06 & 18.5 \\
SL~556   & 0.004 & 0.1 & 1.9 & 1.556 & 0.06 & 18.4 \\
         &       & 0.2 & 2.1 & 1.520 & 0.06 & 18.4 \\
         &       & 0.3 & 2.4 & 1.485 & 0.06 & 18.4 \\
NGC~2155 & 0.004 & 0.1 & 2.7 & 1.396 & 0.03 & 18.5 \\
         &       & 0.2 & 2.9 & 1.377 & 0.03 & 18.5 \\
\hline
\end{tabular}
\end{center}
\end{table}

\clearpage
\begin{table}
\begin{center}
\caption[2]{Mass loss determined from RC luminosity with OS=0.2}
\begin{tabular}{lcrrrrr}
\hline
Object  & $M_{V}$ & RGB($M_{\odot})$ & RC($M_{\odot})$ &  ML($M_{\odot}$) & ML ($\eta$=0.5-0.7)\\
\hline
NGC~2155 & 0.2$\pm 0.1$ & 1.377 & 1.262 & 0.115$\pm$ 0.152 & 0.13-0.18\\
SL~556   & 0.4$\pm 0.1$ & 1.520 & 0.937 & 0.583$\pm$ 0.167 & 0.11-0.16\\
\hline
\end{tabular}
\end{center}
\end{table}

\end{document}